\documentclass[letterpaper, 10 pt, conference]{ieeeconf}  
\usepackage{amsmath,amssymb,amsfonts}
\usepackage{commath}
\usepackage{gensymb}
\usepackage{wasysym}
\usepackage{mathtools}
\usepackage{xfrac}
\usepackage{relsize}
\usepackage{subfig}
\usepackage{siunitx}
\usepackage{multirow}
\usepackage{bm}

\usepackage[usenames]{color}

\DeclareUnicodeCharacter{221E}{\infty}

\newtheorem{remark}{Remark}

\IEEEoverridecommandlockouts                              
\title{\LARGE \bf
Frequency-based Design Method for Model-Free Controllers
}

\author{Marcos Moreno-Gonzalez, Antonio Artuñedo and Jorge Villagra*
\thanks{* Corresponding author.}
\thanks{This work was supported by the Training Program for Academic Staff within the National Plan for Scientific and Technical Research and Innovation by the Ministry of Science, Innovation and Universities of the Government of Spain through the grant referenced FPU22/01412.}
\thanks{M. Moreno-Gonzalez, A. Artuñedo and J. Villagra are with Centre for Automation and Robotics (CAR), CSIC - Universidad Politécnica de Madrid, ctra. de Campo Real, km 0,200, 28500 Arganda del Rey, Spain
        {\tt\footnotesize \{marcos.moreno, antonio.artunedo, jorge.villagra\}@csic.es}}%
}

\begin{document}

\maketitle
\thispagestyle{empty}
\pagestyle{empty}

\begin{abstract}

Model-Free Control (MFC) has been applied to a wide variety of systems in which it has shown its performance. MFC offers ``model-free operation", but the controller design requires some information from the nominal plant. This paper introduces a new design method for model-free controllers that uses minimal data about the system and retrieves a set of stable controller configurations. This method is specifically developed for first-order model-free controllers, but can be extended to second-order controllers, and it relies in a frequency analysis of the controller and the plant. The main feature of the design method is decoupling the design of the main control parameter alpha from the rest, providing specific values for it. The efficacy of the proposed method will be showcased with some relevant application examples.

\end{abstract}

\section{INTRODUCTION}

Model-Free Control (MFC) \cite{fliess.ijc13} has been applied to a wide variety of systems that can be complex, time-varying or non-linear, and it has shown its performance and robustness against plant changes (see \cite{fliess2021alternative,villagra.ijvas09,villagra2020model,amasyali2020hierarchical,baciu2019model,artunedo2024lateral,gedouin2011experimental} as examples of the MFC capabilities). It has been shown that MFC offers ``model-free operation", however, the controller design requires some information from the nominal plant. Recent works \cite{Yahagi2022,hegedHus2022design,moreno2023model} propose different solutions for the design problem that rely on time response information or a nominal model of the plant.

In this paper, a new design procedure for regulators under the MFC paradigm that uses minimal data about the system is presented. This algorithm relies on the frequency analysis of the controller and the regulated system, and is specifically developed for first-order model-free controllers, but can be extended to second-order controllers. The main feature of the proposed algorithm is decoupling the design of the parameter $\alpha$, which has been shown to be related to the aggressiveness \cite{moreno2022speed}, \cite{moreno2023speed} and robustness~\cite{li2022revisit}, from the design of the usual PD gains in the MFC architecture, for which it provides a region that contains the stable parameter configurations.

To evaluate the potential of the frequency-based MFC design method, the longitudinal control of an autonomous car is studied, applying a cascade model-free control structure and simulating the performance of the controller in a realistic vehicle simulator.

The rest of the paper is structured as follows. A brief introduction to Model-Free Control is presented in Section~\ref{MFC_princ}. Section~\ref{alpha_desing} details the proposed design method for the parameter $\alpha$. Section~\ref{PD_desing} explains the proposed method for designing the PD parameters. A simplified version of the design method is shown in Section~\ref{Simp_PD_desing}. Two illustrative examples and the results from simulation tests are presented in Sections~\ref{InvPend} and \ref{VehLong}. Finally, concluding remarks and references can be found in the last section.

\section{MODEL-FREE CONTROL WITH ULTRA-LOCAL MODEL} \label{MFC_princ}
According to \cite{fliess.ijc13}, the dynamics of a system that is nonlinear, time-varying or complex to identify can be replaced by an ultra-local model:
\begin{equation}
y^{(n)} = F + \alpha \cdot u 
\label{eq_ultralocal}
\end{equation}
in which a linear relationship is assumed between the input $u$ and the $n$-th time derivative of the output $y$ with a constant ratio $\alpha$, a design parameter, while $F$ absorbs model errors and system disturbances. Note that $n$ defines the order of the ultra-local model; a low order ($1$ or $2$) has shown to be appropriate for controlling the system \cite{fliess.ijc13}-\cite{gedouin2011experimental}.

The control loop is closed by a so-called \textit{intelligent} controller, typically an iPD controller \cite{fliess2021alternative}:
\begin{equation}
u = \frac{1}{\alpha}  \cdot \left(-F + y_r ^{(n)} + K_p \, e  + K_d \, \dot{e}\right) 
\label{eq2}
\end{equation}
where $u$ is the control action, suffix $r$ stands for reference, $e = y_r - y$ is the tracking error and $K_p$ and $K_d$ are the usual PD control gains. $F$ is a variable that must be estimated in real time using an estimator $\hat{F}$. For this purpose, $F$ can be assumed to be constant between consecutive sample times and can be infered from \eqref{eq_ultralocal} as follows: 
\begin{equation}
\hat{F} (t_k) = \hat{y} ^{(n)} (t_k) - \alpha \cdot u (t_{k - 1}) 
\label{eq3}
\end{equation}
where $t_k$ is the current instant and $\hat{y} ^{(n)}$ is the filtered \mbox{$n$-th} time derivative of $y$. More sophisticated estimators for noise reduction can be found in \cite{fliess.ijc13}, \cite{mboup2007} and \cite{sanyal2022discrete}.

The following filtered derivative operator will be applied in the rest of the paper:
\begin{equation}
D (z) = \frac{1}{T_s} \, \frac{1-z^{-1}}{C +(1-C)\cdot z^{-1}}
\label{eq_filter}
\end{equation}
where $T_s$ is the sample time of the regulator and $C$ is a filtering parameter, which can be experimentally designed for the objective signal so that the measurement noise of the output is reduced.

\begin{remark}\label{remark1}
Note that the error dynamics derived from \eqref{eq_ultralocal} and \eqref{eq2} can be expressed as $e^{(n)}+K_d \dot{e}+K_p e=\hat{F}-F$. If the estimation of $F$ is good enough ($\hat{F} \approx F$), then the system dynamics can be made asymptotically stable through an appropriate choice of the control parameters and order of the ultra-local model.
\end{remark}
%

\section{DESIGN OF $\alpha$} \label{alpha_desing}

\begin{figure}[htbp]
    \centering
    \includegraphics[width=0.85\linewidth]{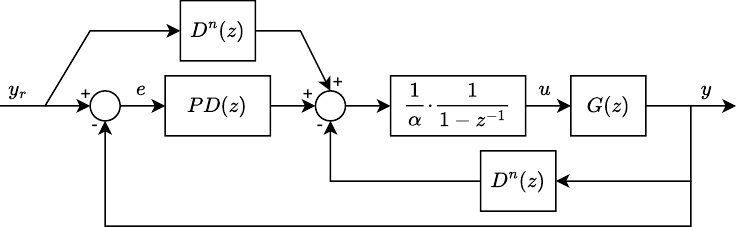}
    \caption{Block diagram of the iPD controller}
    \label{fig:iPD}
\end{figure}

First, considering \eqref{eq2}--\eqref{eq_filter}, the block diagram of a discrete iPD controller controlling a system $G(z)$ can be constructed as shown in Fig. \ref{fig:iPD}. As can be seen in the Figure, the iPD controller can be studied as an Inner Loop Feedback Compensator (ILFC) \cite{fadali2012digital}, an special case of cascade control, when the feedforward term is not considered.

\begin{figure}[!ht]
    \centering
    \includegraphics[width=0.95\linewidth]{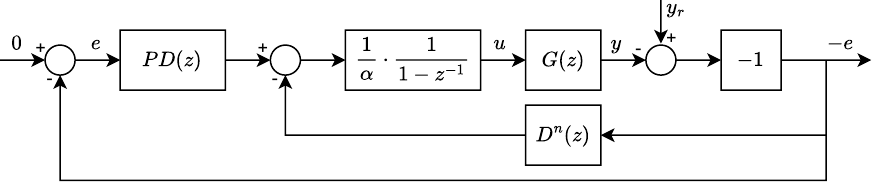}
    \caption{The iPD controller as an ILFC}
    \label{fig:iPD_ceroin}
\end{figure}

From the iPD block diagram in Fig.~\ref{fig:iPD}, the controller can be seen as an Inner Loop Feedback Compensator without neglecting the feedforward filtered derivator $D(z)$. To that end, the feedback loop must be closed by the tracking error~$e$ instead of the system output $y$, as Fig.~\ref{fig:iPD_ceroin} shows. In this case, the system output reference $y_r$ appears as a disturbance and the new reference is $0$. This structure appears naturally in systems that are modeled in error terms, such as the bicycle model for lateral vehicle control \cite{rajamani2011vehicle}, or in systems that are designed for disturbance rejection rather than reference tracking (regulatory operation instead of servo operation), such as the inverted pendulum.

Seeing the iPD as an ILFC leads to an inner loop with (i)~a proportional gain $1/\alpha$, (ii) an integrative pole $\sfrac{1}{1-z^{-1}}$, (iii) the $n$-th filtered derivative $D^n(z)$, and (iv) the system dynamics. An outer loop is then added with a PD structure:

\begin{equation}
PD(z) = K_p + K_d \cdot D(z)
\label{eq_PD}
\end{equation}

ILFC are typically designed by choosing the feedback compensator so that the system dynamics are negligible \cite{heying2003design}, which, in this case, would yield~$D(z)$ to govern the closed loop dynamics. Conversely, a design procedure that minimizes the effect of~$D(z)$ is proposed. Note that minimizing the effect of~$D(z)$ will also reduce the impact of a poor choice of the filter parameter $C$.

\subsection{First Order Ultra-local Model}
In this case, the inner closed loop transfer function is:
\begin{equation}
M_{IL} (z) = \frac{\frac{1}{\alpha} \cdot \frac{G(z)}{1-z^{-1}}}{1+\frac{1}{\alpha} \cdot \frac{G(z)}{1-z^{-1}} \cdot D(z)}
\label{eq_IL}
\end{equation}

To reduce the effect of~$D(z)$, $\alpha$ is designed to make the inner closed loop approximately match the inner direct loop:
\begin{equation}
M_{IL} (z) \approx \frac{1}{\alpha} \cdot \frac{G(z)}{1-z^{-1}}
\label{eq_ILDL}
\end{equation}

This approximation holds iff the inner open loop transfer function $\frac{1}{\alpha} \!\cdot\! \frac{G(z)}{1-z^{-1}} \!\cdot\! D(z)$ is significantly smaller than $1$ in all the frequency spectrum, which yields the expression for $\alpha$:
\begin{equation}
\alpha \gg \abs{ \frac{1}{T_s} \cdot \frac{G\left(e^{i \omega T_s}\right)}{C +(1-C) e^{-i \omega T_s}} } \qquad \forall \omega
\label{eq_alpha1}
\end{equation}

An upper bound of this expression is:
\begin{equation}
\alpha \gg \frac{1}{T_s} \cdot \max \limits_{\omega}{\left\{ \abs{G \left( e^{i \omega T_s} \right) } \right\}} 
\label{eq_alpha1_simp}
\end{equation}

Note that if \eqref{eq_alpha1_simp} holds, \eqref{eq_alpha1} holds as well; since $\alpha$ being much greater than an upper bound, namely \eqref{eq_alpha1_simp}, implies being much greater than the whole expression \eqref{eq_alpha1}. In the rest of the paper, ten times (an order of magnitude) is used as convention for ``much greater".

\subsection{Second Order Ultra-local Model}
In this case, the inner closed loop transfer function is:
\begin{equation}
M_{IL} (z) = \frac{\frac{1}{\alpha} \cdot \frac{G(z)}{1-z^{-1}}}{1+\frac{1}{\alpha} \cdot \frac{G(z)}{1-z^{-1}} \cdot D^2 (z)}
\label{eq_IL2}
\end{equation}

Applying the same procedure as in the previous subsection yields the expression for $\alpha$:
\begin{equation}
\alpha \gg \abs{ \frac{1}{{T_s}^2} \cdot \frac{G\left(e^{i \omega T_s}\right)}{\left(C +(1-C) e^{-i \omega T_s}\right)^2}} \qquad \forall \omega
\label{eq_alpha2}
\end{equation}

An upper bound of \eqref{eq_alpha2} is:
\begin{equation}
\alpha \gg \frac{2}{{T_s}^2} \cdot \max\limits_{\omega}{\left\{\abs{G\left(e^{i \omega T_s}\right)} \right\}} 
\label{eq_alpha1_simp2}
\end{equation}

\section{DESIGN OF THE $PD(z)$ CONTROLLER} \label{PD_desing}

To design $K_p$ and $K_d$, two conditions are considered:
\begin{align}
\abs{\:\!iPD\!\left(\!e^{i \omega T_s}\!\right) \! \cdot G\!\left(\!e^{i \omega T_s}\!\right)\:\!} & < \! 1 &\forall \omega & \in (\omega_{0}, \omega_{N}) \label{eq_module}
\\ 
\mathlarger{\mathlarger{\varangle}} \left\{\!iPD\!\left(\!e^{i \omega T_s}\!\right) \! \cdot G\!\left(\!e^{i \omega T_s}\!\right)\!\right\}\! & >\! -\pi & \forall \omega & < \omega_{0} \label{eq_phase}
\end{align}
where $\omega_{N} = \sfrac{\pi}{T_s}$ is the Nyquist frequency of the system.

Note that this conditions are necessary for the closed loop system to have gain and phase margins, ensuring stability.

\subsection{Module Condition}

While \eqref{eq_ILDL} holds, i.e., $\alpha$ is designed applying \eqref{eq_alpha1_simp} or \eqref{eq_alpha1_simp2} as appropriate, \eqref{eq_module} can be simplified as follows, independently of the order of the ultra-local model:
\begin{equation}
\abs{\tfrac{\left(K_p + K_d \cdot D\left(e^{i \omega T_s}\right)\right) \cdot G\left(e^{i \omega T_s}\right)}{\alpha \cdot \left(1-e^{-i \omega T_s}\right)}} < 1 \quad \forall \omega \in (\omega_{0}, \omega_{N})
\label{eq_module_simp}
\end{equation}

Substituting $D(z)$ from \eqref{eq_filter}, yields:
\begin{multline}
    \abs{\tfrac{\left(K_p T_s C + K_d - \left( K_p T_s (C-1) + K_d\right)e^{-i \omega T_s}\right)}{\alpha T_s \cdot \left(C+(1-C) \cdot e^{-i \omega T_s}\right) \cdot \left(1 -e^{-i \omega T_s}\right)}} \! \cdot \abs{G\left(e^{i \omega T_s}\right)} < 1 \\
    \forall \omega \in (\omega_{0}, \omega_{N})
\label{eq_module_simp1}
\end{multline}

Expanding the left hand side of the module:
\begin{multline}
\tfrac{\sqrt{ K_p'^2 \left(2CC'+1\right)+2K_p' K_d C''+2K_d^2 - 2\left(K_p'^2CC'+K_p' K_d C''+K_d^2\right) \cdot B}}{\sqrt{\left(C^2+C'^2-2CC'\cdot B\right) \cdot \left(2-2B\right)}} \\
 < \tfrac{\alpha T_s}{\abs{G\left(e^{i \omega T_s}\right)}} \quad \forall \omega \in (\omega_{0}, \omega_{N})
\label{eq_module_simp2}
\end{multline}
where
\begin{align*}
C' &= C - 1; \quad C'' = 2C - 1 \\
K_p' &= K_p \cdot T_s; \quad B = \cos{\left(\omega T_s\right)}
\end{align*}

This expression defines a set of elliptical shapes in the \mbox{$K_pK_d$-plane} as a function of~$\omega$, in which the smallest one is a boundary of the stability region. For systems whose module decreases at high frequencies, i.e., systems that act as low pass filters, the smaller ellipse from \eqref{eq_module_simp2} appears when $\omega = \omega_0$, being $\omega_0$ the phase crossover frequency of $\sfrac{G(z)}{C +(1-C) z^{-1}}$.

\subsection{Phase Condition}

The left hand side of \eqref{eq_phase} can be expanded as follows:
\begin{multline}
\mathlarger{\mathlarger{\varangle}} \!\left(K_p T_s C \!+\! K_d \!+\! 1 - \left( K_p T_s (C\!-\!1) \!+\! K_d \!+\! 1\right)e^{-i \omega T_s}\right) \\ 
+ \mathlarger{\mathlarger{\varangle}} \, \tfrac{G \left(e^{i \omega T_s}\right)}{\left(C+(1-C) \cdot e^{-i \omega T_s}\right) \cdot\left(1 -e^{-i \omega T_s}\right)} > -\pi \;\;\;\; \forall \omega < \omega_{0}
\label{eq_phase_simp1}
\end{multline}

Obtaining the phase with the $\arctan$ and simplifying with trigonometric properties, the condition can be rewritten as:
\begin{multline}
\tfrac{\sin{\omega T_s}}{\frac{K_p T_s C + K_d + 1}{K_p T_s (C-1) + K_d + 1} - \cos{\omega T_s}} > \\
- \tan{ \mathlarger{\mathlarger{\varangle}} \biggl\{ \!\tfrac{G \left(e^{i \omega T_s}\right)}{\left(C+(1-C) \cdot e^{-i \omega T_s}\right) \cdot\left(1 -e^{-i \omega T_s}\right)} \!\biggr\}} \;\;\;\; \forall \omega < \omega_{0}
\label{eq_phase_simp2}
\end{multline}

Further simplification of this inequation is not straightforward since the sign of the trigonometric functions is variable and thus the inequality orientation could change. However, it is found that taking \eqref{eq_phase_simp2} as an equality, it defines the following set of straight lines in the \mbox{$K_pK_d$-plane} as a function of $\omega$:
\begin{align}
K_d & = K_p T_s \frac{\left( C-W(C-1) \right)}{W-1} - 1 \label{eq_phase_simp3} \\
W & = \cos{\omega T_s} - \tfrac{\sin{\omega T_s}}{\tan{ \mathlarger{\mathlarger{\varangle}} \biggl\{ \! \frac{G \left(e^{i \omega T_s}\right)}{\left(C+(1-C) \cdot e^{-i \omega T_s}\right) \cdot\left(1 -e^{-i \omega T_s}\right)} \!\biggr\}}} \label{eq_phase_simp4}
\end{align}

Notice that the line obtained when taking $\omega = \sfrac{\omega_0}{2}$ is a bound to the stability set.

Eq. \eqref{eq_phase_simp3} has been obtained for first order model-free controllers because the assumption made by~\eqref{eq_ILDL} holds for the module but does not have to be true for the phase.

In summary, the proposed design method requires the following information from the plant: (i) its maximum magnitude for \eqref{eq_alpha1_simp2}, and (ii) its phase response to obtain $\omega_0$ and for \eqref{eq_phase_simp4}.

\section{SIMPLIFIED DESIGN METHOD} \label{Simp_PD_desing}

A simplified version of the design method can be obtained when the controlled plant information is scarce. 

\subsection{Simplified module condition}

Eq. \eqref{eq_module_simp2} can be approximated in two ways:

\begin{enumerate}
\item Substituting $\abs{G\left(e^{i \omega T_s}\right)}$ by $\max\limits_{\omega}{ \left\{\abs{G\left(e^{i \omega T_s}\right)} \right\}}$. This leads to a conservative approximation, i.e., a stabilizing set smaller than the real one. This approximation is useful for stable systems.
\item Substituting $\omega$ in $\abs{G\left(e^{i \omega T_s}\right)}$ by the phase crossover frequency of $G(z)$, $\omega_1$. This leads to a permissive approximation, i.e., a stabilizing set bigger than the real one.
\end{enumerate}

\subsection{Simplified phase condition}

When little plant information is available, another phase condition can be obtained from the frequency analysis of the controller. Therefore, the controller $(PD(z))$ needs at least to increase the phase of the closed-loop system:
\begin{equation}
    \mathlarger{\mathlarger{\varangle}} \tfrac{K_p T_s C \!+\! K_d + 1 - \left( K_p T_s (C-1) + K_d + 1\right)e^{-i \omega T_s}}{1 -e^{-i \omega T_s}} > 0
\end{equation}
which yields the following relation between $K_p$ and $K_d$:
\begin{equation}
2 (K_d + 1) > -K_p T_s (2C-1) 
\label{eq_phase_simp_10}
\end{equation}

This expression is a necessary (not sufficient) condition for stability, i.e., it is the lowest bound of the stability set for every system.

In summary, with these approximations, the proposed design method requires less information from the plant: (i) its maximum magnitude for \eqref{eq_alpha1_simp2}, as the design of $\alpha$ is not further simplified, and (ii) optionally its phase crossover frequency $\omega_1$.

\section{INVERTED PENDULUM CONTROL DESIGN} \label{InvPend}

The design method is initially tested in an academic example, for which the inverted pendulum is adequate since it is an unstable system with strong non-linearities.

\begin{figure}[htbp]
    \centering
    \includegraphics[width=0.35\linewidth]{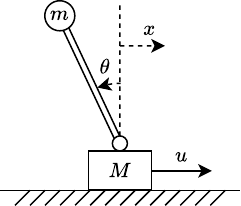}
    \caption{Inverted pendulum with cart}
    \label{fig:inv_pend}
\end{figure}

The inverted pendulum is modeled from Fig. \ref{fig:inv_pend} as follows:
\begin{align}
(M+m) \cdot \ddot{x} &= u - ml\ddot{\theta} \label{eq_inv_pend1} \\
(I+ml^2) \cdot \ddot{\theta} &= ml \ddot{x} + mgl\theta - b \dot{\theta} \label{eq_inv_pend2}
\end{align}

\noindent where $u$ is the force applied to the cart, $x$ is the longitudinal displacement of the cart, $\theta$ is the angle of the pendulum, $M = 0.1\,\si{kg}$ and $m = 0.5\,\si{kg}$ are the cart and pendulum masses, $l = 0.5\,\si{m}$ is the length of the pendulum, $I=m\cdot l^2$ is the inertia of the pendulum weight, $b = 2\,\si{kg\cdot m^2/s}$ is the viscous friction of the pendulum with the cart and $g = 9.8\,\si{m/s^2}$ is the acceleration due to gravity. The aim is to control the angle of the pendulum, therefore, $\theta$ is obtained from \eqref{eq_inv_pend1} and \eqref{eq_inv_pend2} as:
\begin{equation}
    \left(I+ml^2+\frac{m^2 l^2}{M+m}\right) \cdot \ddot{\theta} + b \dot{\theta} - mgl\theta = \frac{m l}{M+m} u
    \label{eq_inv_pend}
\end{equation}

\begin{figure}[htbp]
    \centering
    \includegraphics[width=0.9\linewidth]{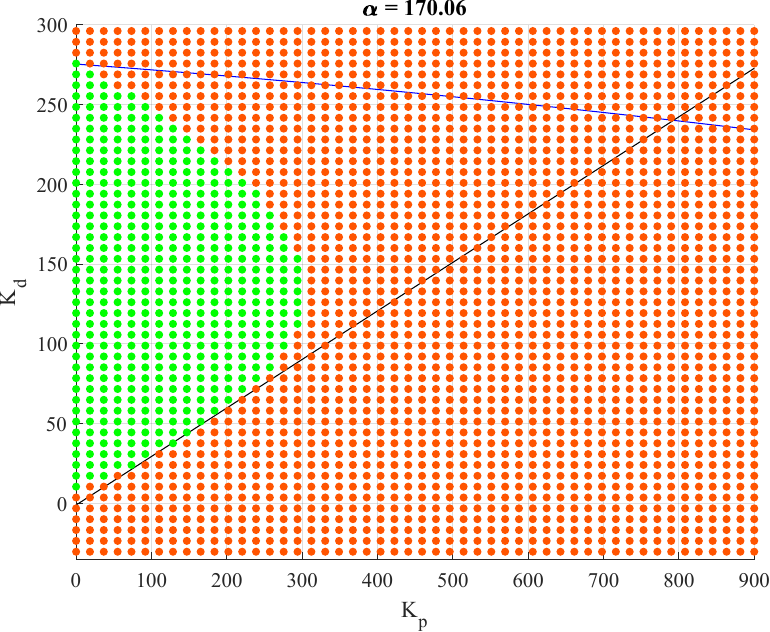}
    \caption{Stability set of the inverted pendulum controller}
    \label{fig:inv_pend_stability_set}
\end{figure}

This system is then transformed into a discrete transfer function with a sample time $T_s = 0.01\,\si{s}$. If the derivative filter $D(z)$ is designed with $C=4$, the proposed MFC design method is applied using \eqref{eq_alpha1_simp}, resulting in $\alpha$ needing to be greater than $17.006$; in order to set a fixed value for the next design steps, the value increased by an order of magnitude is taken. The design procedure result is presented in Fig.~\ref{fig:inv_pend_stability_set}, where the module and phase conditions have been applied, obtaining the blue and black lines respectively, which are bounds of the stabilizing set. The control configurations inside the resulting set are evaluated and depicted in green if they result in a stable closed loop, and red otherwise. As the system is originally unstable, the method provides a set of configurations bigger than the actual stabilizing set.

\begin{figure}[htbp]
    \centering
    \includegraphics[width=0.9\linewidth]{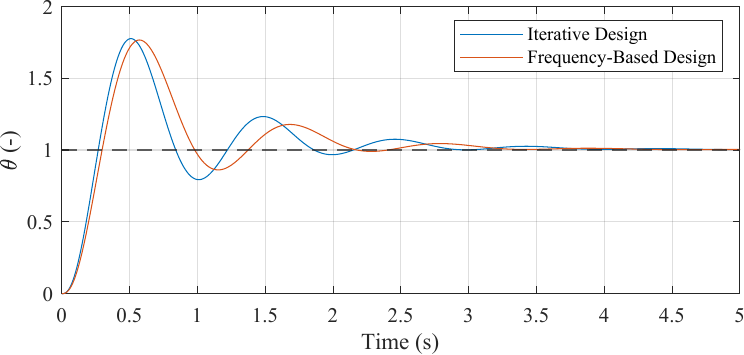}
    \caption{Step response of the inverted pendulum}
    \label{fig:inv_pend_step}
\end{figure}

Configurations in Fig.~\ref{fig:inv_pend_stability_set} are tested and that with the best Integral of the Absolute Error~($IAE$) is chosen, getting the following control parameters: $\alpha=170.06$, $K_p=48.98$, $K_d=64.92$. In parallel, an iterative optimization process is performed to design an alternative iPD controller, obtaining the configuration \mbox{($\alpha=154.94$, $K_p=48.56$, $K_d=71.05$)} that minimizes the $IAE$. The step response of both controllers is plotted in Fig. \ref{fig:inv_pend_step}.

As can be seen in Fig. \ref{fig:inv_pend_step}, the step response of the iteratively designed controller is similar to the response of the controller designed with the proposed method. This shows that the capacity of the proposed method of decoupling the design of $\alpha$ from the PD gains does not significantly affect the controller performance, while requiring less information from the plant.

\section{VEHICLE LONGITUDINAL CONTROL DESIGN} \label{VehLong}

The design method is also tested for a real application: the longitudinal control of an autonomous vehicle.

\subsection{Identified vehicle longitudinal dynamics}

The longitudinal dynamics of the vehicle have been identified from a real car with the aid of MATLAB\copyright 's System Identification Toolbox as the following $\mathcal{Z}$-transfer function with a $97.05\%$ of accuracy w.r.t. the real signal:
\begin{equation}
G(z) = \frac{0.01262 z^{-1} - 0.01236 z^{-2}}{1 - 2.957 z^{-1} + 2.915 z^{-2} - 0.9581 z^{-3}}
\label{eq_ident_long_dynam}
\end{equation}
being the throttle-brake pedals ($u \in [-1,1]$) the input of the system, the speed (\si{m/s}) the output of the system and with a sample time $T_s = 0.05\,\si{s}$, fixed by the GNSS period.

\begin{figure}[htbp]
    \centering
    \includegraphics[width=0.95\linewidth]{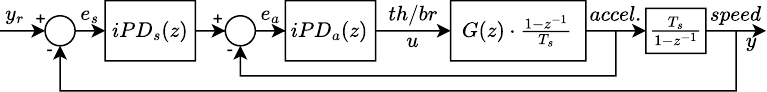}
    \caption{Cascade control structure for the vehicle longitudinal dynamics}
    \label{fig:cascade}
\end{figure}

The controller structure chosen to regulate the plant is a cascade acceleration--speed structure, as shown in Fig. \ref{fig:cascade}.

\subsection{Inner loop iPD design}

The proposed MFC design method is applied to the inner loop system \mbox{$G(z) \cdot \tfrac{1-z^{-1}}{T_s}$}, as the inner loop reflects the acceleration dynamics. Firstly, after studying the signal-to-noise ratio in the real sensor, the derivative filter $D(z)$ is designed with the parameter $C=7.5$. Secondly, using \eqref{eq_alpha1_simp}, $\alpha$ must be greater than $147.63$; in order to set a fixed value, the value increased by an order of magnitude is taken as was done in the previous example. 

\begin{figure}[htbp]
\centering
    \subfloat[Complete method]{
        \includegraphics[width=0.87\linewidth]{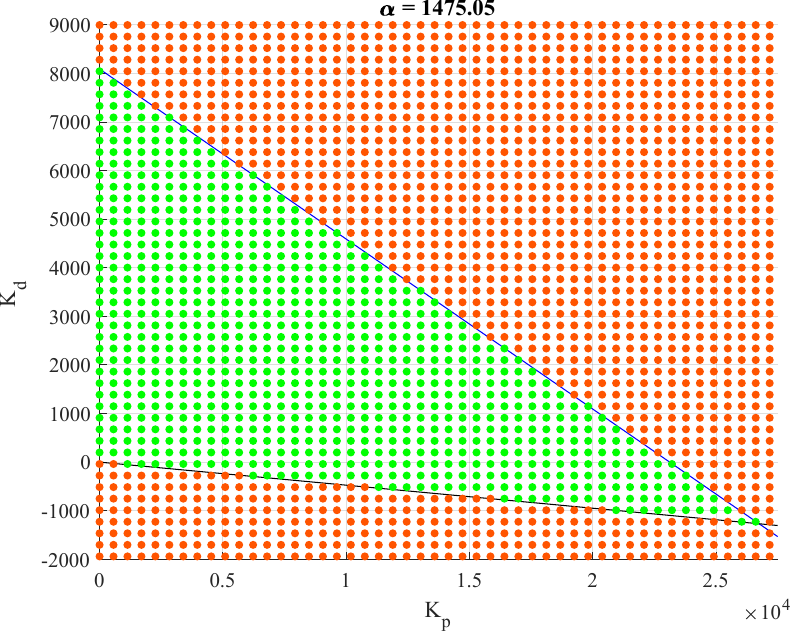}
        \label{fig:inner_set_compl}}

    \subfloat[Simplified method]{
        \includegraphics[width=0.87\linewidth]{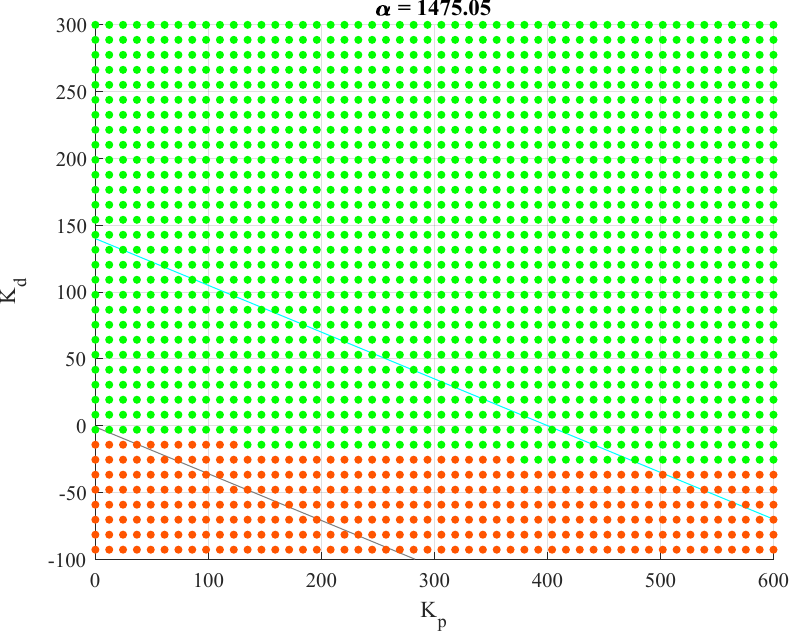}
        \label{fig:inner_set_simp}}
\caption{Stability set of the inner controller}
\label{fig:inner_set}
\end{figure}

Fig.~\ref{fig:inner_set_compl} shows the control configuration set obtained when applying the module and phase conditions (in blue and black respectively) of the complete design procedure. Configurations inside and around the resulting sets are evaluated, finding that all the insiders are stable (in green in the Figure). The results obtained when the simplified version of the design method is applied are plotted in Fig.~\ref{fig:inner_set_simp}, where the module and phase conditions are the cyan and gray lines respectively. As can be seen in the Figure, the simplified module condition applied defines a much smaller area than the complete version, and the simplified phase condition is more permissive than the complete one.

Choosing the most appropriate configuration is still not straightforward; however, \cite{villagra2020model} shows that iP-iP cascade configurations (where $K_d = 0$) have similar characteristics as the well known P-PI cascade structure. Besides, it is found that smoother inner responses (using a low $K_p$) are easier to stabilize by the outer controller. In this case, the conservative approximation applied in the simplified design procedure (in Fig.~\ref{fig:inner_set_simp}), can be useful for choosing $K_p$.

With these considerations, the inner controller parameters are set as: $\alpha = 1475.05, K_p = 20, K_d = 0$.

\subsection{Outer loop iPD design}

The design procedure is applied again taking the system as the closed inner loop plus the integrator. First, the derivative filter parameter is set to $C=3.5$ to filter the noise from the real sensor. Secondly, from \eqref{eq_alpha1_simp}, $\alpha \gg 15864.4$ is obtained. 

\begin{figure}[htbp]
    \centering
    \includegraphics[width=0.9\linewidth]{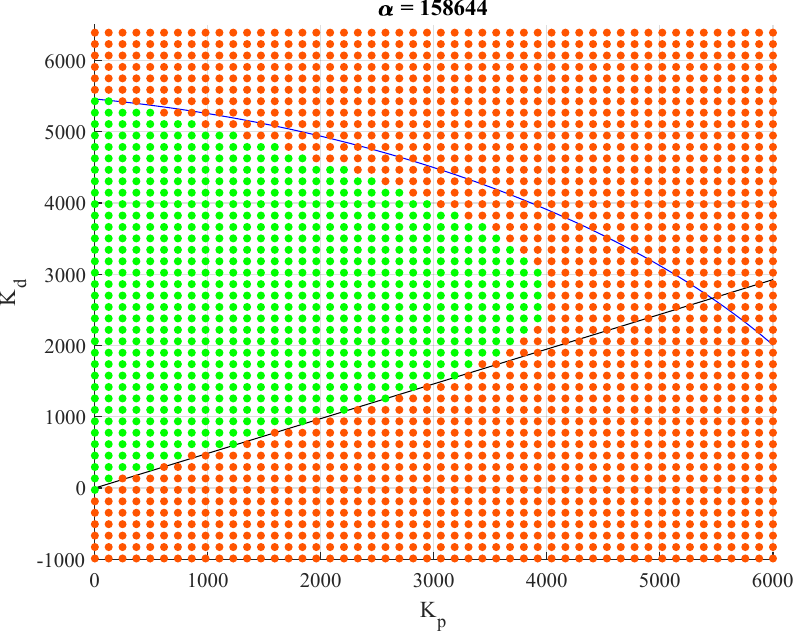}
    \caption{Stability set of the outer controller}
    \label{fig:outer_set}
\end{figure}

Last, the module and phase conditions retrieve the blue and black lines respectively that define the configuration set in Fig. \ref{fig:outer_set}. Stability of the control configurations is evaluated and plotted in the Figure in green if stable and red if unstable. As the system is more complex, the method does not provide the stability set precisely, but the region obtained encompasses the stability set.

Simulations have been carried out in order to test control configurations inside the resulting set. In these tests, it has been found that, for the outer loop, a controller with low $K_p$ and medium $K_d$ works best, as it results in a smooth behavior which can follow the varying speed references with little oscillation. Consequently, the outer controller parameters are set as $\alpha = 158644$, $K_p = 3$, $K_d = 3000$.

\subsection{Evaluation in a realistic simulator}

The model of the vehicle used in simulation tests in this section was aimed to mimic an experimental platform (described in \cite{artunedo2019decision}) with high precision. The longitudinal dynamics of the model, which is the control objective, include the following parts: 

\begin{enumerate}
\item A dynamic model with 14 degrees of freedom (6 for the vehicle body motion: longitudinal, lateral, vertical, roll, pitch, and yaw; and 8 for the wheels: vertical motion and spin of each wheel).

\item A power-train model, which comprises: (i)~a torque map of the engine identified from measurements taken in the experimental platform; (ii)~the gearbox, that includes the drive ratios and gear shifting logic of the vehicle; (iii)~resistive forces from the braking system, wind and gravity.

\item The tire response is characterized with the Pacejka tire model~\cite{Pacejka1992}.
\end{enumerate}

\begin{figure}[htbp]
    \centering
    \includegraphics[width=0.95\linewidth]{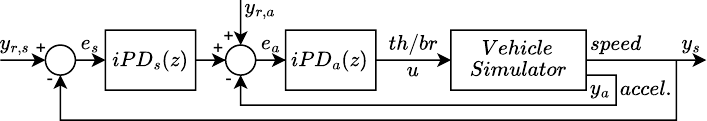}
    \caption{Real cascade control structure for the vehicle longitudinal control}
    \label{fig:cascade_real}
\end{figure}

It has been found that adding the acceleration reference, as Fig.~\ref{fig:cascade_real} shows, provides better results specially when the speed is high, as it allows the controller to react faster when the speed reference changes suddenly. A preview point is applied in this reference, which is adaptive with the speed: $d_p = PT \cdot v_x$, where $d_p$ is the preview distance from the center of gravity of the vehicle, $v_x$ is the longitudinal speed of the vehicle and $PT$ is the constant ratio that relates both, for which~$0.6$ has been found to perform best with the designed controller. Apart, a controller designed iteratively is tested. The parameters of both controllers used in the tests are gathered in Table \ref{table_control_params}.

\begin{table}
\caption{Control parameters used in simulation tests}
\label{table_control_params}
\begin{center}
\begin{tabular}{|c|c|c|c|c|c|c|} \hline
\textbf{Controller} & \textbf{Controller} & $\bm{C}$ & $\bm{\alpha}$ & $\bm{K_p}$ & $\bm{K_d}$ & $\!\bm{PT}$\! \\ \hline
\multirow{2}{*}{\begin{tabular}[c]{@{}c@{}}\!Freq-Based\! \\Design\end{tabular}} & Outer (speed)  & 3.5   & 158644     & 3       & \!3000\!    & \multirow{2}{*}{\!0.6\!}  \\ \cline{2-6}
                                                                             & Inner (accel.) & 7.5   & \!1475.05\!    & 20      & 0       &                       \\ \hline
\multirow{2}{*}{\begin{tabular}[c]{@{}c@{}}Iterative\\Design\end{tabular}}   & Outer (speed)  & 2.5   & 330     & \!$10^{-4}$\!       & \!3.655\!    & \multirow{2}{*}{\!0.7\!}  \\ \cline{2-6}
                                                                             & Inner (accel.) & 9.6   & 2063    & \!28.13\!      & 0       &                       \\ \hline
\end{tabular}
\end{center}
\end{table}

\begin{figure}[htbp]
\centering
    \subfloat[Speed reference and vehicle's speed]{
        \includegraphics[width=0.9\linewidth]{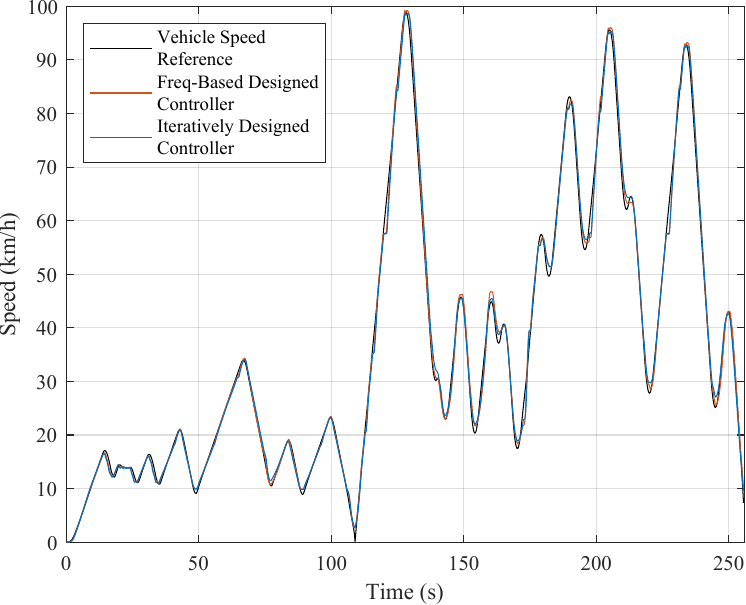}
        \label{fig:response_speed}}
        
    \subfloat[Speed error over time]{
        \includegraphics[width=0.9\linewidth]{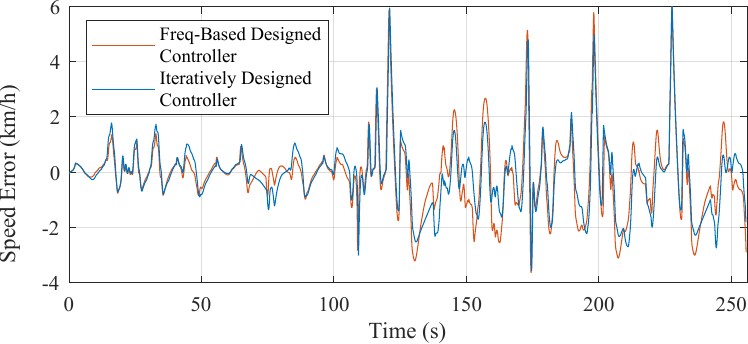}
        \label{fig:response_error}}

    \subfloat[Pedal opening over time]{
        \includegraphics[width=0.9\linewidth]{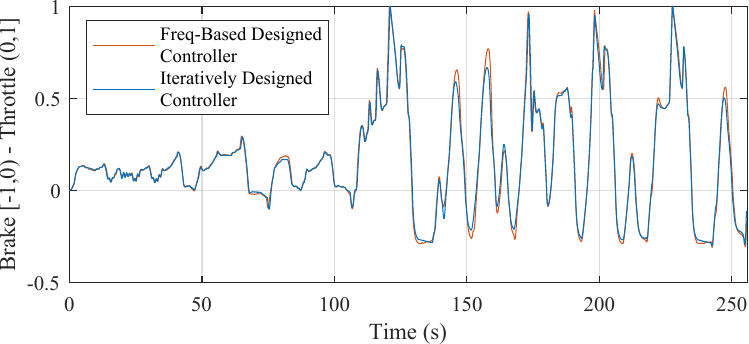}
        \label{fig:response_pedal}}
\caption{Response of the designed controllers in the realistic simulator}
\label{fig:response}
\end{figure}

The controllers designed have been tested with the speed profile of a slow urban circuit concatenated to a fast highway path, so that it presents an uniform distribution of longitudinal speeds up to $100\,\si{km/h}$. The response of the controllers following such trajectory in the aforementioned vehicle simulator is shown in Fig. \ref{fig:response}.

As can be seen in Fig. \ref{fig:response_speed}, the reference speed profile covers low and high speed scenarios, with noticeable speed changes, where both controllers are able to show their capacity to regulate the vehicle with a reasonable error. The error in the first part of the profile, i.e., the urban trajectory, is within $\pm1.7\,\si{km/h}$, while in the second part is higher, reaching up to $6\,\si{km/h}$, as can be observed in Fig. \ref{fig:response_error}. Note that error peaks appear for both controllers in the same occasions, i.e., at gear shifting (positive peaks) and during reference transitions from acceleration to braking (negative peaks). 

The vehicle is a complex system in which metrics regarding comfort and safety need to be evaluated in addition to the performance metric assessed in the previous example. Therefore, response of the controllers is evaluated in:
\begin{enumerate}
    \item Integral of the Absolute tracking Error ($IAE$), which reflects the general tracking quality of the controller.
    \item Integral of the Absolute value of the second Derivative of the Control Action ($IAUDD$), which reflects the abrupt changes in the control action.
    \item Overshooting ($OS$), which reflects the responsiveness to reference changes while rejects unavoidable errors such as those when shifting gears. Also, the overshooting considers only negative errors, as being faster than the reference can cause safety issues.
\end{enumerate}

\begin{table}[!ht]
\caption{Performance metrics obtained in Simulation Tests}
\label{table_performance_metrics}
\begin{center}
\begin{tabular}{|c|c|c|c|} \hline
\textbf{Controller}        & $\!\bm{IAE}\,(\si{km/h})\!$ & $\!\bm{IAUDD}\,(-)\!$ & $\!\bm{OS}\,(\si{km/h})\!$  \\ \hline
Freq-Based Design & 0.8620             & 0.2323       & 3.6333             \\ \hline
Iterative Design  & 0.8347             & 0.2207       & 3.5520             \\ \hline
\end{tabular}
\end{center}
\end{table}

The metric values obtained, gathered in Table \ref{table_performance_metrics}, show that both controllers exhibit a high tracking quality, with less than $1\,\si{km/h}$ of $IAE$, good responsiveness, with less than $4\,\si{km/h}$ of $OS$, while keeping a smooth control action, with less than $0.25$ of $IAUDD$. As can be seen in the Table, the iteratively designed controller barely outperforms the frequency-based designed controller; however, the iterative design requires a complete model of the system.

\section{CONCLUSIONS AND FUTURE WORKS}

The aim of this work was to develop a design method for Model-Free Controllers requiring minimal information about the controlled system. The proposed method, which relies in a frequency analysis of the controller and the plant, requires little information about the system, and a version that requires minimal information is also included. The method retrieves a set of stable controller configurations and is specifically developed for first-order model-free controllers, but can be extended to second-order controllers. The main feature of the method is to decouple the design of the main control parameter alpha from the rest.

The proposed MFC design method has been successfully applied to the design of an inverted pendulum controller and a longitudinal controller for an autonomous vehicle.

Future works will explore the addition of target stability margins, namely gain and phase margins, as design requirements for the MFC design method.




\bibliographystyle{IEEEtran}
\bibliography{IEEEabrv,main}             
                                                     
\end{document}